\documentclass[journal,comsoc]{IEEEtran}
\usepackage[T1]{fontenc}
\usepackage{times}
\usepackage{soul}
\usepackage{url}
\usepackage[hidelinks]{hyperref}
\usepackage[utf8]{inputenc}
\usepackage[small]{caption}
\usepackage{graphicx}
\usepackage{amsmath}
\usepackage{amsthm}
\usepackage{booktabs}
\usepackage{algorithm}
\usepackage{algorithmic}
\usepackage{multirow}
\usepackage{threeparttable}
\usepackage{amsfonts}
\usepackage{float}
\usepackage{url}
\usepackage{bm}
\usepackage{xcolor}
\usepackage{subfigure}
\usepackage{amssymb}
\usepackage{makecell}
\usepackage{bbding}

\newcommand{\cmmnt}[1]{}

\ifCLASSINFOpdf
\else
\fi
\usepackage{amsmath}
\usepackage[cmintegrals]{newtxmath}
\begin{document}
%
\title{DiCLET-TTS: Diffusion Model based Cross-lingual Emotion Transfer for Text-to-Speech --- A Study between English and Mandarin}
\author{Tao Li,
        Chenxu Hu,
        Jian Cong,
        Xinfa Zhu,
        Jingbei Li,
        Qiao Tian,
        Yuping Wang,
        Lei Xie,~\IEEEmembership{Senior Member,~IEEE}
\vspace{-0.4cm}
 \thanks{This work was supported by the National Key Research and Development Program of China under Grant 2020AAA0108600. (Corresponding author: Lei Xie)}

\thanks{Tao Li, Xinfa Zhu, and Lei Xie is with the School of Computer Science, Northwestern Polytechnical University, Xi’an 710072, China. Email: taoli@npu-aslp.org (Tao Li), xfzhu@mail.nwpu.edu.cn (Xinfa Zhu), lxie@nwpu.edu.cn (Lei Xie)}

\thanks{Chenxu Hu is with the Institute for Interdisciplinary Information Sciences, Tsinghua University, Beijing 100084, China, Email: chenxuhu65@gmail.com}
\thanks{Jingbei Li is with the Department of Computer Science and Technology, Tsinghua University, Beijing 100084, China, Email: lijb19@mails.tsinghua.edu.cn}
\thanks{Jian Cong, Qiao Tian, and Yuping Wang are with the Speech, Audio, and Music Intelligence (SAMI) Group, ByteDance, Shanghai 200233, China. Email: congjian.tts@bytedance.com (Jian Cong), tianqiao.wave@bytedance.com (Qiao Tian), wangyuping@bytedance.com (Yuping Wang)}
    }


%

%

\markboth{Journal of \LaTeX\ Class Files,~Vol.~14, No.~8, August~2015}%
{Wang \MakeLowercase{\textit{et al.}}: Bare Demo of IEEEtran.cls for IEEE Communications Society Journals}
%



\maketitle

\begin{abstract}

While the performance of cross-lingual TTS based on monolingual corpora has been significantly improved recently, generating cross-lingual speech still suffers from the foreign accent problem, leading to limited naturalness. Besides, current cross-lingual methods ignore modeling emotion, which is indispensable paralinguistic information in speech delivery.  
In this paper, we propose DiCLET-TTS, a Diffusion model based Cross-Lingual Emotion Transfer method that can transfer emotion from a source speaker to the intra- and cross-lingual target speakers.
Specifically, to relieve the foreign accent problem while improving the emotion expressiveness, the terminal distribution of the forward diffusion process is parameterized into a speaker-irrelevant but emotion-related linguistic prior by a prior text encoder with the emotion embedding as a condition.
To address the weaker emotional expressiveness problem caused by speaker disentanglement in emotion embedding, a novel orthogonal projection based emotion disentangling module (OP-EDM) is proposed to learn the speaker-irrelevant but emotion-discriminative embedding. 
Moreover, a condition-enhanced DPM decoder is introduced to strengthen the modeling ability of the speaker and the emotion in the reverse diffusion process to further improve emotion expressiveness in speech delivery.
Cross-lingual emotion transfer experiments show the superiority of DiCLET-TTS over various competitive models and the good design of OP-EDM in learning speaker-irrelevant but emotion-discriminative embedding.

\end{abstract}


%
\IEEEpeerreviewmaketitle

\vspace{-0.2cm}
\section{Introduction}
\label{sc:Introduction}

\IEEEPARstart{C}{ross}-lingual text-to-speech (TTS)~\cite{zhang2019learning,rallabandi2017building,cai2023cross} refers to the task that requires the system to generate speech in a language foreign to a target speaker. This task has many applications, such as code-mixed speech synthesis for a voice agent, foreign movie dubbing~\cite{hu2021neural}, and computer-assisted pronunciation teaching~\cite{pourhosein2020using}.
Due to the difficulty of obtaining a bilingual corpus produced by a highly proficient speaker in both languages, more practically, current studies mainly build a cross-lingual TTS system based on corpora from monolingual speakers in different languages~\cite{9414226,xin2020cross,shang2021incorporating,ye2022improving}.
However, these approaches mostly ignore modeling emotion aspects during speech generation, while emotion is a kind of indispensable paralinguistic information that reveals the speaker's intentions and moods. 
Without properly delivering such paralinguistic information, the gap between synthetic and real speech cannot be mitigated.
This paper aims to address this emotional speech synthesis problem in cross-lingual TTS with only monolingual corpora available. 
Specifically, a \textit{cross-lingual emotion transfer} method in the same-gender scenario is introduced.
With cross-lingual emotion transfer, a cross-lingual TTS model can directly synthesize emotionally diverse speech in a language foreign to the target speaker, i.e., synthesizing emotional speech in authentic Mandarin for an English speaker by transferring the emotion from a Mandarin speaker, without employing any Mandarin emotional speech from the English speaker during system building. 

Although the current studies have made many efforts to cross-lingual TTS, there is still a gap between generated speech and those of native speakers in terms of naturalness, as synthetic speech often comes with a strong \textit{foreign accent}~\cite{zhang2019learning}.
The reason for this phenomenon is that each speaker in the training set speaks only one language, and the entanglement between different speech factors, such as linguistic content, speaker identity, and emotion, makes it hard to only transfer the speaker's timbre across different languages.
Therefore, the key to alleviating this foreign accent issue is how to properly \textit{disentangle} the speaker and language or linguistic content~\cite{shang2021incorporating,ye2022improving,zhan22_interspeech}.

Based on the disentanglement strategy, the existing cross-lingual approaches can be roughly divided into \textit{implicit-based} and \textit{explicit-based} methods~\cite{ye2022improving}.
Implicit-based methods mainly study the unified linguistic/phonetic representations across languages to disentangle language and speaker timbre implicitly~\cite{9053094,zhao2020towards,li2019bytes,bansal2020improving,maniati2021cross,zhan2021improve}. 
On the other hand, to further solve the foreign accent problem, the explicit-based methods prefer to adopt adversarial learning~\cite{zhang2019learning,xin2020cross,ye2022improving,nekvinda2020one} or mutual information~\cite{9414226} to minimize the correlation between different speech factors, thus encouraging the model to automatically learn disentangled linguistic representations.
However, the disturbance caused by adversarial learning could degrade the naturalness of the generated cross-lingual speech. 
Furthermore, the above cross-lingual studies have not considered the emotion factor yet, while proper emotion is essential to speech expressiveness, as just mentioned.

To improve the emotion diversity of synthetic cross-lingual speech, we need to implement a cross-lingual TTS model with the ability of cross-speaker emotion transfer as well, which can produce emotional speech for target speakers by transferring the emotion from another source speaker~\cite{Li2021ControllableCE}.
Reference-based style transfer is the most popular strategy for cross-speaker emotion transfer, where Reference Encoder~\cite{Skerry2018Towards}, Global Style Tokens (GST)~\cite{Kwon2019EmotionalSS}, and Variational Auto-Encoder (VAE)~\cite{Zhang2019LearningLR} are typically used to extract an emotion embedding from the reference Mel-spectrum with desired emotion.
Usually, the speaker identity can be obtained from either a trainable look-up table~\cite{Gibiansky2017DeepV2} or a pre-trained speaker verification model~\cite{Jia2018TransferLF}. 

The key to the reference-based methods is to learn speaker-irrelevant emotion embedding from the reference spectrum by disentangling the emotion and the speaker's timbre~\cite{Bian2019Multi,2019Multi,Li2021ControllableCE}.
Otherwise, the speaker information retained in the emotion embedding could contaminate the target speaker's timbre, making synthesized speech sound somehow like uttered by the source speaker rather than the target speaker, i.e., the \textit{speaker leakage} problem~\cite{Karlapati2020CopyCatMF}. 
However, due to the emotion and the timbre being deeply entangled in speech, it is hard to remove the speaker-related information while avoiding the emotion information from being weakened in the emotion embedding, which could lead to \textit{weaker emotional expressiveness} problem in the synthesized speech~\cite{li22h_interspeech}. 
Furthermore, for cross-lingual emotion transfer, a unique challenge is that emotion will make the intonation change more violently~\cite{levis1999intonation} and then aggravate the influence of foreign accents, resulting in a serious decline in the naturalness of the emotional speech synthesized for foreign speakers.
In this paper, we attempt to enable English speakers to express various emotions in Mandarin naturally and expressively, which is a more challenging scenario since Mandarin is a typical tonal language and English is a non-tonal language~\cite{duanmu2004tone,li2021human}.

Recently, diffusion probabilistic models (DPMs)~\cite{sohl2015deep,song2020score} have shown their superiority in various content generation tasks~\cite{song2020denoising,rombach2022high,kawar2022denoising}, including the recent attempts in speech generation tasks~\cite{popov2021grad,chen2021wavegrad,jeong2021diff,huang2022fastdiff}.
A DPM aims to gradually transform the raw data into a terminal distribution (usually standard Gaussian) by a forward diffusion process and then learns a reverse diffusion process parameterized with a neural network to rebuild the raw data from the terminal distribution~\cite{sohl2015deep}. 
Importantly, DPMs show superiority in expressive data generation, which means they can generate more diverse data due to their ability to essentially preserve the semantic structure of the data.
To leverage the advances of DPMs, this paper proposes \textbf{DiCLET-TTS}, a novel DPM-based TTS model for cross-lingual emotion transfer.
DiCLET-TTS consists of a prior text encoder, an orthogonal projection based emotion disentangling module (OP-EDM), and a condition-enhanced DPM decoder. 

Specifically, to relieve the \textit{foreign accent} problem and improve emotional expressiveness, the prior text encoder aims to parameterize the terminal distribution of the forward diffusion process into a speaker-irrelevant but emotion-related linguistic prior, achieved by two steps.
First, the linguistic encoding is constrained by speaker adversarial training to obtain a speaker-irrelevant linguistic representation. 
A content loss is particularly adopted to mitigate the interference of adversarial training on linguistic encoding.
An emotional adaptor is subsequently adopted to convert the speaker-irrelevant linguistic representation into a speaker-irrelevant but emotion-related linguistic prior with the condition of emotion embedding extracted from OP-EDM. 

To address the \textit{weaker emotional expressiveness} problem, the emotion embedding space learned in OP-EDM is explicitly constrained by an Orthogonal Projection Loss~\cite{ranasinghe2021orthogonal} to force the emotion embeddings to be aggregated within the same emotion category and orthogonal between different emotion categories, leading to a discriminative emotion embedding space and improved transferred emotion expressiveness in synthetic speech.

The reverse diffusion process is further parameterized with the DPM decoder to restore the target Mel-spectrum from the speaker-irrelevant but emotion-related terminal distribution. 
We particularly introduce a \textit{condition-enhanced} decoder to further improve emotion expressiveness in speech delivery. 
Specifically, the decoder follows the Unet~\cite{ronneberger2015u} structure in Grad-TTS~\cite{popov2021grad}, but differently, the speaker and emotion embeddings are fed to each ResBlock as \textit{enhanced conditions}.

During the experimental evaluation, different emotions are transferred from the source speaker to the intra- and cross-lingual target speakers, respectively, to verify the effectiveness of DiCLET-TTS while comparing the performance difference between intra- and cross-lingual emotion transfer. 
Results show that although the performance of intra-lingual transfer is better than that of more challenging cross-lingual transfer, DiCLET-TTS can clearly improve speech naturalness, emotion similarity, and speaker similarity compared to three competitive methods in both intra- and cross-lingual emotion transfer scenarios. 
Furthermore, the embedding visualization and preference test demonstrates the advantages of OP-EDM in learning speaker-irrelevant but emotion-discriminative embedding.

The rest of this paper is organized as follows. 
Section~\ref{sc:related work} reviews the related work. 
Section~\ref{sc:method} introduces the proposed method in detail.  
Section \ref{sc:experiments} and Section \ref{sc:results} describe the experimental setups and results, respectively.
The component analysis is introduced in Section~\ref{component}. 
Finally, the paper concludes in Section \ref{sc:conclusion}. 
Examples of synthesized speech can be found on the project page\footnote{The demo can be found on https://silyfox.github.io/DiCLETdemo/ \label{ft:homepage}}.

\section{Related work}
\label{sc:related work}
This section describes related studies on cross-lingual, cross-speaker emotion transfer, and recent DPM-based TTS. 

\vspace{-0.2cm}
\subsection{Cross-lingual TTS}

Most current studies realize cross-lingual TTS by mixing monolingual corpora of different languages while disentangling the speaker and language or linguistic representations in implicit or explicit ways to alleviate the foreign accent problem.
Implicit methods mainly focus on exploring language-irrelevant input representations~\cite{9053094,zhao2020towards,bansal2020improving,maniati2021cross}.
Liu et al.~\cite{liu2020multi} introduce a shared phoneme set for different languages. Language embedding is extended by tone/stress embeddings to control the accent of synthetic speech.
In~\cite{9053094,zhao2020towards}, the Automatic Speech Recognition (ASR) models are employed to extract language-irrelevant Phonetic Posterior Gram (PPG) features as the input representations. 
Unicode bytes~\cite{li2019bytes}, mixed-lingual Grapheme-to-Phoneme (G2P)~\cite{bansal2020improving} frontend, and International Phonetic Alphabet (IPA)~\cite{chen2019cross,maniati2021cross,zhan2021improve} are also taken as the unified phonetic representations that share pronunciation across languages~\cite{ye2022improving}.
These studies indicate that language-irrelevant representations can help disentangle speaker and language, but the complexity of the cross-lingual TTS pipeline is increased.

The explicit methods encourage the cross-lingual model to automatically learn disentangled representation, i.e., speaker-irrelevant linguistic representations or language-irrelevant speaker representations.
Zhang et al.~\cite{zhang2019learning} and Nekvinda et al.~\cite{nekvinda2020one} employ domain adversarial training to remove speaker identity entangled in linguistic representations.
Xin et al.~\cite{xin2020cross} construct a language-irrelevant speaker space via domain adaptation and perceptual similarity regression. 
In~\cite{9414226}, mutual information minimization and domain adversarial training are adopted to disentangle the obtained language and speaker embedding, which guides cross-lingual speech synthesis.
Ye et al.~\cite{ye2022improving} introduce a triplet training scheme to enhance cross-lingual pronunciation by allowing previously unseen content and speaker combinations to be seen during training.    
Shang et al.~\cite{shang2021incorporating} alleviate the foreign accent problem by using existing authentic style during inference and accordingly propose a style encoder through adversarial training.
The above studies mainly address the foreign accent problem, while emotional speech is not considered.

\vspace{-0.1cm}
\subsection{Cross-speaker emotion transfer}

Cross-speaker emotion transfer in TTS shares similar methods with other kinds of style transfer, as emotions are expressed in a special style. 
For clarity, all of them are referred to as emotion transfer.
Currently, there are mainly two major approaches for cross-speaker emotion transfer, i.e., label-assisted and reference-based methods. 
Label-assisted\cite{pan21d_interspeech,xie2021multi} methods are proposed to predict emotion-related prosodic information, i.e., pitch and energy, from input text with speaker and emotion ID. 
However, since prosodic information contained in text lack residual acoustic information other than pitch and energy, these methods are prone to produce synthesized speech with average expressiveness.

The reference-based methods~\cite{Skerry2018Towards,Wang2018Style,Zhang2019LearningLR,zhang2022iemotts,Li2021ControllableET,liu2021expressive} is the mainstream strategy, which learns an emotion representation~\cite{liu22i_interspeech} from reference as a condition to guide emotion transfer.
Skerry-Ryan et al.~\cite{Skerry2018Towards} integrate the Tacotron~\cite{wang2017tacotron,shen2018natural} model with an extra prosody encoder, denoted as Reference Encoder, in which the reference is encapsulated into a fixed-length embedding that is directly concatenated with the linguistic representations.
Global Style Tokens (GST)~\cite{Wang2018Style} further extends Reference Encoder by an embedded library to learn a latent high-dimensional representation.
Variational Auto-Encoder~\cite{Zhang2019LearningLR} is also introduced to learn the potential emotion representation from the reference to complete emotion transfer. 
However, these methods ignore speaker disentanglement and aggregate all emotion-related aspects, e.g., pitch, energy, and speaker's timbre, into one hidden emotion embedding, resulting in speaker leakage.
To achieve speaker disentanglement, Bian et al.~\cite{Bian2019Multi} propose a multi-reference encoder and an intercross training scheme in which emotion and speaker are disentangled and transferred independently.
Whitehill et al.~\cite{2019Multi} improve the performance of the multi-reference model on disjoint datasets by unpaired training strategy and adversarial cycle consistency scheme.
Li et al.~\cite{Li2021ControllableCE} introduce an emotion disentangling module, which constrains the emotion embedding to be speaker-irrelevant via an orthogonal loss with the learned speaker embedding.
To summarize, the aforementioned methods mainly aim to obtain a speaker-irrelevant emotion embedding space in different ways, while the trade-off between speaker timbre and emotional expressiveness is inevitable~\cite{Karlapati2020CopyCatMF,li22h_interspeech}.

\vspace{-0.2cm}
\subsection{DPM-based TTS}

The Diffusion Probabilistic Models (DPMs) aim to convert the raw data distribution into random noise before reversing the transformations step by step to rebuild a new sample with the same distribution as the raw data~\cite{sohl2015deep,cao2022survey} and have achieved the SOTA results in various tasks, e.g., image generation~\cite{dhariwal2021diffusion,song2020denoising}, super-resolution~\cite{kawar2022denoising,rombach2022high}, and TTS~\cite{popov2021grad,chen2021wavegrad,jeong2021diff,popov2021diffusion}. 
One major drawback of DPM-based models is the slow sampling speed due to many iterative steps. 
Therefore, many previous DPM-based TTS methods focus on accelerating the sampling method to boost the inference speed~\cite{jeong2021diff,huang2022fastdiff,leng2022binauralgrad,huang2022prodiff,yang2022diffsound}.
Some research considers changing the training process to generate high-quality speech. 
Grad-TTS~\cite{popov2021grad} and PriorGrad~\cite{lee2021priorgrad} transform the raw data distribution into a data-dependent prior distribution obtained from the conditional information. 
The studies~\cite{kim2022guided,kim22guided} also drive unconditional DPM-based models trained on untranscribed speech to generate high-quality samples by phoneme classifier guidance, where the phoneme classifier is trained separately.
Liu et al.~\cite{liu2022diffsinger} and Xue et al.~\cite{xue22c_interspeech} introduce the DPM-based model into singing voice synthesis (SVS), demonstrating its superiority in the expressiveness synthesis tasks.
In this study, we introduce the DPM-based model to cross-lingual emotion transfer TTS, a more challenging and unexplored task.

\begin{figure*}[htb]	
	\centering
	\includegraphics[width=18cm]{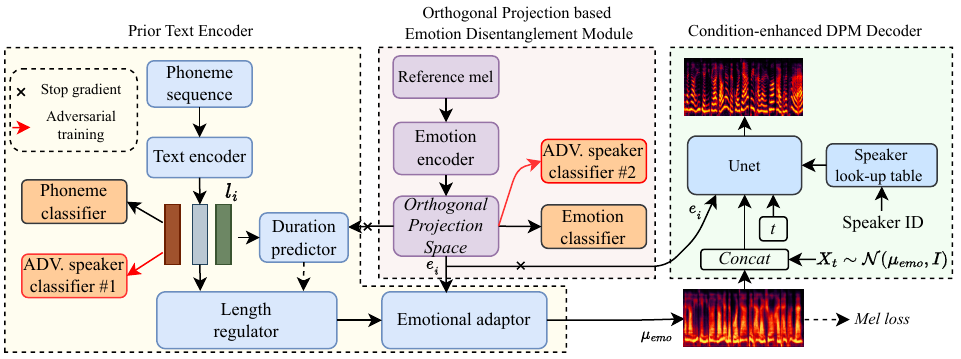} 
	\caption{The architecture of the proposed DiCLET-TTS. The input text is represented as the phoneme sequence, and speech is represented by Mel-spectrum, which can be converted to the waveform by a Hifi-Gan vocoder.}
	\label{fig:frame1}
\vspace{-0.2cm} 
\end{figure*}

\section{Methodology}
\label{sc:method}

This section first gives a system overview of the proposed DiCLET-TTS and then introduces the design of each module in detail.

Figure~\ref{fig:frame1} illustrates the proposed DiCLET-TTS architecture for cross-lingual emotion transfer, a DPM-based TTS model consisting of three major components: a prior text encoder, an orthogonal projection based emotion disentangling module (OP-EDM), and a condition-enhanced DPM decoder. 
As discussed, the entangled speaker and linguistic representation can lead to a \textit{foreign accent} problem.
Thus, the prior text encoder is adopted to parameterize the forward diffusion's terminal distribution as a speaker-irrelevant but emotion-related linguistic prior, to mitigating the foreign accent while improving emotional expression.
The speaker identity is only modeled by a speaker look-up table with speaker ID in the DPM decoder to further disentangle the speaker from other factors.
Considering that the speaker disentanglement in emotion embedding could lead to \textit{weaker emotional expressiveness} in synthesized speech, our disentangled emotion embedding space is further constrained by an introduced orthogonal projection loss to ensure that the embedding maintains intense emotion discrimination after removing speaker-related information.
Finally, a condition-enhanced DPM decoder is adopted to restore the target Mel-spectrum from the speaker-irrelevant but emotion-related terminal distribution, guided by the speaker and emotion embeddings.

\vspace{-0.3cm}
\subsection{Prior text encoder}
\label{sc:text_encoder}

The prior text encoder consists of a text encoder, a length regulator, and an emotional adaptor, aiming to parameterize the terminal distribution of the forward diffusion process into a speaker-irrelevant but emotion-related linguistic prior. 
Specifically, to remove the speaker information from the linguistic representation $l_i$, the text encoder is encouraged to encode input phonemes in a speaker-irrelevant manner by introducing a speaker adversarial classifier.
Then, we encode the length-regulated $l_i$ through an emotional adaptor under the condition of emotion embedding (will be introduced in Section~\ref{sc:edm}) to obtain the speaker-irrelevant but emotion-related linguistic representation $\mu_{emo}$.
However, adversarial training could disturb linguistic encoding to some extent. 
Therefore, a content loss is introduced to mitigate this disturbance.
Details of the prior text encoder and the loss functions will be introduced.

\subsubsection{Text encoder}
The text encoder converts the phoneme sequence into the hidden linguistic representation $l_i \in \mathbb{R}^{C_i \times d}$, where $C_i$ denotes the length of the phoneme sequence, and $d$ denotes the dimension of representation.
The speaker adversarial classifier is to make the linguistic representation $l_i$ speaker-irrelevant through the softmax layer with gradient reversal (GR), and the loss function is defined as:

\begin{equation}
    {{\cal L}_{ladv}} =  - \sum\limits_{i = 1}^n {\log } P\left( {{s_i}\mid {l_i}} \right),
\end{equation}
where $n$ is the batch size, $P( {{s_i}\mid {l_i}})$ is possibility of $l_i$ belonging to the speaker $s_i$.
So that we can minimize the speaker classification loss to reversely optimize the text encoder on the speaker classification task.

The content loss guarantees the text encoder's stability to encode the input phoneme sequence when using the speaker adversarial classifier.
The corresponding loss function is defined as:
\begin{equation}
    {{\cal L}_{c}} =  - \sum\limits_{i = 1}^n \sum\limits_{j = 1}^{C_i} {\log } P\left( {{p_i^{j}}\mid {l_i^{j}}} \right).
\end{equation} 
where $p_i^{j}$ denotes the ground-truth label of the $j$-th phoneme in the $i$-th input sequence, and $l_i^{j}$ denotes the $j$-th hidden linguistic representation of the $i$-th input sequence.

\subsubsection{Length regulator}
The length regulator has the same architecture as that in FastSpeech~\cite{Ren2019FastSpeechFR}.
It takes emotion embedding as an extra input since the duration of the same sentence in different emotions should be different. 
The ${l_i}$ is length-regulated according to its real duration by the length regulator during training.
The duration predictor is trained by the mean square error (MSE) loss with the ground-truth duration. The duration loss is denoted as $\mathcal{L}_{dur}$. 

\subsubsection{Emotional adaptor}
The emotional adaptor aims to transform the length-regulated ${l_i}$ into a speaker-irrelevant but emotion-related linguistic representation $\mu_{emo}$ through multiple FFT blocks with Conditional LayerNorm~\cite{chen2020adaspeech}, which takes the emotion embedding as the condition. The $\mu_{emo}$ has the same dimension as the Mel-spectrum and is adopted to define the forward diffusion's terminal distribution ($\mathcal{N}(\mu_{emo}, I)$.
An MSE loss $\mathcal{L}_{mel}$ constrains the $\mu_{emo}$ from the target Mel-spectrum.
Regarding the prior text encoder, the total objective function is defined as:
\begin{equation}
{{\cal L}_{prior}} = 0.01*{{\cal L}_{ladv}} + {{\cal L}_{c}} + {{\cal L}_{dur}} + {{\cal L}_{mel}}. 
\end{equation}

\vspace{-0.4cm}
\subsection{Orthogonal projection based emotion disentanglement module}
\label{sc:edm}

The orthogonal projection based emotion disentanglement module (OP-EDM) is to learn an emotion encoder that extracts the speaker-irrelevant emotion embedding $e_i$ from the reference. Ideally, the embedding $e_i$ should be free of speaker-related information and discriminative in distinguishing different emotion categories. 
To this end, the emotion encoder in OP-EDM is trained with two loss functions: 1) an adversarial loss to make the obtained embedding $e_i$ speaker-irrelevant; 2) a classification loss to make the obtained embedding $e_i$ emotion-dependent.
Specifically, the emotion encoder in OP-EDM has a similar architecture as the Reference Encoder~\cite{Skerry2018Towards} to generate a 256-dimensional vector as the emotion embedding $e_i$.

The adversarial loss aims to make the emotion embedding $e_i$ speaker indistinguishable. 
A GRL is adopted between the emotion encoder and a speaker classifier.
Then, the emotion encoder is reversely optimized on the speaker classification by minimizing the following loss function:
\begin{equation}
    {{\cal L}_{sadv}} =  - \sum\limits_{i = 1}^n {\log } P\left( {{s_i}\mid {e_i}} \right),
\end{equation}
where $P( {{s_i}\mid {e_i}})$ is the possibility of the emotion embedding $e_i$ extracted from speech with the speaker label $s_i$.

The classification loss is implemented by an emotion classifier with the same structure as the above speaker classifier, to make the obtained $e_i$ emotion-dependent.
Note that the sentences of all non-emotional speakers are treated as a separate emotion category, denoted as \textit{neutral\_N}.
Thus, the softmax layer produces the probability of $8$ emotion types, i.e., $neutral$, $happy$, $surprise$, $angry$, $disgust$, $fear$, $sad$, and $neutral\_N$. The corresponding objective function is: 

\begin{equation}
    {{\cal L}_{emo}} =  - \sum\limits_{i = 1}^n {\log } P\left( {{t_i}\mid {e_i}} \right),
\end{equation}
where $P( {{t_i}\mid {e_i}})$ is possibility of emotion embedding $e_i$ belonging to the emotion label $t_i$.

\subsubsection{Explicit constraint for emotion embedding space}
As mentioned, the emotional information conveyed by the $e_i$ would be weakened after removing the speaker-related information, leading to weaker emotional expressiveness in synthesized speech. 
To address this issue, we resort to the Orthogonal Projection Loss~\cite{ranasinghe2021orthogonal} (OPL), a potent technique to construct discriminative embedding space without learnable parameters.
The objective of OPL is to enforce constraints to embedding space such that the embedding $e_i$ for different emotion classes $t_i$ is orthogonal to each other and the $e_i$ for the same class is similar, which can effectively disentangle the class-specific characteristics of different emotions, further improving the emotion discrimination of $e_i$. 
The objective function is defined as: 
\begin{equation}
\mathcal{L}_{opl}=(1-E_{same})+0.5 * |E_{different}|,
\end{equation}
where $|\cdot|$ is the absolute value operator. 
When minimizing this loss $\mathcal{L}_{opl}$, the first term $(1-E_{same})$ can ensure clustering of same class samples, while the second term $|E_{different}|$ can ensure the orthogonality of different class samples.
The $E_{same}$ and $E_{different}$ are defined as:

\begin{equation}
E_{same}=\sum_{\substack{t_{i}=t_{j}}}^n\left\langle\mathbf{e}_{i}, \mathbf{e}_{j}\right\rangle, \quad
E_{different}=\sum_{\substack{t_{i} \neq t_{k}}}^n\left\langle\mathbf{e}_{i}, \mathbf{e}_{k}\right\rangle,
\end{equation}
where $\langle\cdot, \cdot\rangle$ is the cosine similarity operator applied on two emotion embeddings. 
The total objective function of OP-EDM is defined as:
\begin{equation}
{{\cal L}_{op-edm}} = 0.2*{{\cal L}_{sadv}} + 0.8*{{\cal L}_{emo}} + {{\cal L}_{opl}}. 
\end{equation}

\subsection{Condition-enhanced DPM decoder}

A DPM with data-dependent prior can be seen as such: a forward diffusion converts the raw data into simple terminal distribution (usually standard Gaussian) by gradually adding Gaussian noise, then based on this terminal distribution, a reverse diffusion parameterized with a neural network learns to follow the trajectories of the reverse-time forward diffusion~\cite{song2020score,popov2021diffusion}.
If the forward and reverse diffusion processes have close trajectories, then the distribution of generated samples will be very close to that of the raw data.

\begin{table*}[h]
\centering
\caption{Dataset for the cross-lingual emotion transfer TTS.}
\setlength{\tabcolsep}{3.0mm}
\label{tab:data}
\begin{tabular}{c|c|c|ccccccc|c}
\toprule
\multirow{2}{*}{Corpus} & \multirow{2}{*}{Gender} & \multirow{2}{*}{Language} & \multicolumn{7}{c|}{Emotion (sentences)}   & \multirow{2}{*}{ Usage} \\
 &    &    & Neutral & Happy  & Surprise &Sadness & Angry & Disgust & Fear &   \\ \midrule
CN1     & Female   & Mandarin  &5k   &-    &-     &-    &-   &-   &-       &Training\&Evaluation   \\
CN2     & Female   & Mandarin  &5k    &-   &-     &-    &-   &-   &-       &Training               \\
CN-emo  & Female   & Mandarin  &5k   &2k    &2k    &2k   &2k  &2k  &2k     &Training\&Evaluation   \\
EN1     & Female   & English  &9k   &-   &-     &-    &-   &-   &-        &Training\&Evaluation   \\ 
EN2     & Female   & English  &9k   &-    &-     &-    &-   &-   &-       &Training   \\ \bottomrule
\end{tabular}
\end{table*}

In DiCLET-TTS, the terminal distribution of forward diffusion has been parameterized by the prior text encoder as a simple linguistic-based prior distribution $\mathcal{N}(\mu_{emo}, I)$, which is emotion-related but speaker-irrelevant. 
We parameterize the reverse diffusion with a condition-enhanced DPM decoder to further improve the emotion expressiveness in speech delivery.
Specifically, the condition-enhanced DPM decoder's architecture is based on Unet and is the same as that in Grad-TTS~\cite{popov2021grad}, but the speaker and emotion embeddings are added to each ResBlock rather than just concatenated with the decoder's input. 
The speaker and emotion embeddings are produced by a speaker look-up table and the OP-EDM, respectively.

We mostly follow the formulation introduced in Grad-TTS, the forward and reverse diffusion processes of DiCLET-TTS as satisfies the following It$\widehat{o}$ stochastic differential equations (SDEs):

\begin{equation}
\begin{aligned}
d X_t=\frac{1}{2} \Sigma^{-1}\left(\mu_{emo}-X_t\right) \beta_t d t+\sqrt{\beta_t} d \overrightarrow{W}_t,
\end{aligned}
\end{equation}

\begin{equation}
\begin{aligned}
\label{eq2}
d X_t=&\left(\frac{1}{2}\Sigma^{-1}\left(\mu_{emo}-X_t\right)-\nabla \log p_t\left(X_t \mid X_0\right)\right) \beta_t d t
\\&+\sqrt{\beta_t} d \overleftarrow{W}_t,
\end{aligned}
\end{equation}
where the $\overrightarrow{W}_t$ and $\overleftarrow{W}_t$ are forward and reverse-time Brownian motion. The $X_{0}$ and $X_{t}$ are raw and noise data, where $X_{t}\sim\mathcal{N}(\mu_{emo}, I)$, $t \in[0,1]$.
The $\beta_{t}$ is a noise schedule with the same definition in Grad-TTS. 
The $\log p_{t}\left(X_{t} \mid X_{0}\right)$ is the log probability density function which is predicted by a learnable score function $s_{\theta}\left(X_{t}, \mu_{emo}, t, E_{spk}, e_{i}\right)$ parameterized with the condition-enhanced DPM decoder $\theta$.
The $E_{spk}$ and $e_{i}$ are speaker and emotion embeddings, respectively.
The reverse diffusion (\ref{eq2}) is solved by a defined ordinary differential equation (ODE):

\begin{equation}
d X_{t}=\frac{1}{2}\left(\mu_{emo}-X_{t}-s_{\theta}\left(X_{t}, \mu_{emo}, t, E_{spk}, e_{i}\right)\right) \beta_{t} d t.
\end{equation}
This reverse diffusion process is trained by minimizing weighted L2 loss as follows:

\begin{equation}
\begin{aligned}
\mathcal{L}(\theta)_{diff}=&\underset{\theta}{\arg \min } \int_{0}^{1} \lambda_{t} \mathbb{E}_{X_{0}, X_{t}}\left\|s_{\theta}\left(X_{t}, \mu_{emo}, t, E_{spk}, e_{i}\right)
\right.\\ &\left.
-\nabla \log p_t\left(X_{t} \mid X_{0}\right)\right\|_{2}^{2} d t,
\end{aligned}
\end{equation}
where the $\lambda_{t} = 1-e^{-\int_{0}^{t} \beta_{s} ds}, 0<s<t$.
In brief, the reverse diffusion parameterized with the $s_{\theta}\left(X_{t}, \mu_{emo}, t, E_{spk}, e_{i}\right)$ is trained to approximate gradient of log-density of $X_{t}$ given $X_{0}$, $E_{spk}$, $e_{i}$ and $\mu_{emo}$. 
During the inference, we first predict a speaker-irrelevant but emotion-related $\mu_{emo}$ from input text with the emotion embedding $e_{i}$.
The $e_{i}$ is extracted by OP-EDM from the reference with desired emotion. 
Then the condition-enhanced DPM decoder gradually reconstructs the target Mel-spectrum using the score predicted from $s_{\theta}$ in adjustable iterations, with the conditions of $e_{i}$ and $E_{spk}$.

\vspace{-0.1cm}
\subsection{Final objective function}
All modules introduced in the previous sections are trained together. 
The final objective function of the proposed DiCLET-TTS is defined as:
\begin{equation}
    {{\cal L}_{total}} ={{\cal L}_{prior}} + {{\cal L}_{op-edm}} + {{\cal L}_{diff}},
\end{equation}
where ${\cal L}_{prior}$, ${\cal L}_{op-edm}$, and ${\cal L}_{diff}$ are loss functions of the prior text encoder, OP-EDM, and condition-enhanced DPM decoder. 

\vspace{-0.1cm}
\section{Experimental Setups}
\label{sc:experiments}

This section introduces the database configuration, evaluation methods, training setups, and compared methods.

\vspace{-0.2cm}
\subsection{Dataset} 
\label{sc:database}

As shown in Table~\ref{tab:data}, the dataset used in this paper comprises five female monolingual speakers, denoted as CN1, CN2, CN-emo, EN1, and EN2.
Note that CN1 and CN2 are publicly available Mandarin corpus\footnote{The dataset is available at \url{http://www.data-baker.com/hc_znv_1.html}} and EN1 and EN2 are internal English corpora.
Only \textbf{CN-emo} is the emotional corpus employed as the \textbf{source speaker} during emotion transfer. 
All data are studio-quality recorded at 48KHz.

The test set consists of $1100$ sentences in total.
Specifically, we randomly select $700$ sentences from the \textit{CN-emo} corpus, and each emotion category (including \textit{neutral}) contains $100$ sentences. 
In addition, $400$ sentences are randomly selected from the four neutral speaker corpora, and every speaker contains $100$ sentences.

\vspace{-0.2cm}
\subsection{Model Configurations}
\label{sc:model_config}

The text encoder has the same architecture in DelightfulTTS~\cite{liu2021delightfultts}, which is composed of a pre-net (3 layers of convolutions followed by a fully-connected layer), 
6 Conformer blocks~\cite{gulati2020conformer} with multi-head self-attention, and the final linear projection layer to generate 448-dimensional linguistic representation. 
The speaker adversarial classifier in the text encoder consists of a GRU layer, a fully connected (FC) layer, and a softmax layer. Especially, a gradient reversal layer (GRL) is adopted between the GRU and the FC layer. 
The content loss is implemented by a phoneme classifier consisting of two FC layers and a softmax layer.
The emotional adaptor consists of 1 layer of 1D convolution, 2 FFT blocks, and a 1D convolution output layer, where each FFT block is followed by a Conditional LayerNorm~\cite{chen2020adaspeech}.
We employ the same structure for speaker and emotion classifiers in OP-EDM: an FC layer and a softmax layer. 
The difference is that a GRL layer is inserted before the speaker classifier.

\vspace{-0.3cm}
\subsection{Evaluation Methods} 
\label{sc:database}

Three types of human perceptual rating experiments are performed:
1) Mean Opinion Score (MOS)~\cite{shang2021incorporating} is used for subjective evaluation of the naturalness, which can reflect the influence of foreign accents and emotion on synthesized naturalness.
2) Differential Mean Opinion Scores (DMOS)~\cite{Li2021ControllableCE} is adopted to subjectively evaluate the synthesized speech from two aspects, emotion similarity (between the synthesized speech and compared emotional reference) and speaker similarity (between the synthesized speech and compared target speaker's reference).
3) AB preference test~\cite{an2022disentangling} (AB test) is adopted to compare samples synthesized by two models, where participants are asked to choose which speech sample sounds closer to the compared reference in terms of speaker or emotion. 
In both MOS and DMOS tests, the participants are asked to rate given speech a score ranging from $1$ to $5$ based on the specific purpose.
The rating criteria is: \textit{bad = 1}; \textit{poor = 2}; \textit{fair = 3}; \textit{good = 4}; \textit{great = 5}, in 0.5 point increments.

\begin{table*}[t]
\caption{Naturalness MOS results of DiCLET-TTS with M3, CSET, and Grad-TTS in transferring emotion to the intra- and cross-lingual target speakers, with confidence intervals of 95$\%$. The bold indicates the best performance of the four models in each emotion.}
\label{tab:naturalness}
\setlength{\tabcolsep}{2.8mm}
\centering
\begin{tabular}{l|c|cccc|cccc}
\toprule
\multirow{2}{*}{Emotion} & \multirow{2}{*}{Language} & \multicolumn{4}{c|}{Intra-lingual scenario (target Mandarin speaker)}       & \multicolumn{4}{c}{Cross-lingual scenario (target English speaker)}         \\ \cmidrule{3-10}
                         &                           & \multicolumn{1}{c}{M3} & \multicolumn{1}{c}{CSET} & \multicolumn{1}{c}{Grad-TTS} & \multicolumn{1}{c|}{DiCLET-TTS} & \multicolumn{1}{c}{M3} & \multicolumn{1}{c}{CSET} & \multicolumn{1}{c}{Grad-TTS} & \multicolumn{1}{c}{DiCLET-TTS} \\ \midrule
\multirow{2}{*}{Neutral}   &  Mandarin                     & 4.17$\pm$0.03 & 4.15$\pm$0.05 & 4.19$\pm$0.04 & \bf{4.23}$\pm$0.06 & 3.92$\pm$0.06 & 3.81$\pm$0.04 & 3.87$\pm$0.06      & \bf{3.98}$\pm$0.05  \\
                           &  English                     & 3.95$\pm$0.04 & 3.84$\pm$0.02 & 3.88$\pm$0.07 & \bf{3.99}$\pm$0.05 & 4.18$\pm$0.04 & 4.14$\pm$0.06 & \bf{4.24}$\pm$0.03 & 4.21$\pm$0.05       \\
\midrule
Fear      &  \multirow{6}{*}{Mandarin}    & 4.05$\pm$0.05 & 3.92$\pm$0.08 & 4.03$\pm$0.04 & \bf{4.07}$\pm$0.08 & 3.82$\pm$0.05 & 3.51$\pm$0.07 & 3.68$\pm$0.04      & \bf{3.90}$\pm$0.06  \\
Disgust   &                              & 4.08$\pm$0.09 & 4.05$\pm$0.10 & 4.10$\pm$0.09 & \bf{4.12}$\pm$0.07 & 3.87$\pm$0.08 & 3.69$\pm$0.09 & 3.72$\pm$0.09      & \bf{3.93}$\pm$0.04  \\
Angry     &                              & 4.00$\pm$0.10 & 3.93$\pm$0.07 & 4.02$\pm$0.05 & \bf{4.03}$\pm$0.05 & 3.76$\pm$0.03 & 3.42$\pm$0.10 & 3.59$\pm$0.10      & \bf{3.82}$\pm$0.07  \\
Sadness   &                              & 4.03$\pm$0.09 & 3.99$\pm$0.04 & 4.04$\pm$0.09 & \bf{4.06}$\pm$0.08 & 3.81$\pm$0.06 & 3.57$\pm$0.08 & 3.69$\pm$0.07      & \bf{3.88}$\pm$0.07  \\
Happy     &                              & 4.01$\pm$0.04 & 3.98$\pm$0.05 & 4.00$\pm$0.07 & \bf{4.04}$\pm$0.03 & 3.75$\pm$0.09 & 3.46$\pm$0.06 & 3.61$\pm$0.08      & \bf{3.84}$\pm$0.08  \\
Surprise  &                              & 4.02$\pm$0.07 & 3.96$\pm$0.06 & 4.05$\pm$0.04 & \bf{4.09}$\pm$0.02 & 3.73$\pm$0.07 & 3.44$\pm$0.11 & 3.64$\pm$0.08      & \bf{3.83}$\pm$0.05  \\
\bottomrule
\end{tabular}
\end{table*}

During our experiments, we found that the results of different speakers in the same language were similar in human evaluation.
Consequently, to reduce the cost of human evaluation, we randomly selected one speaker from each of the two languages, i.e., CN1 and EN1, as our target speakers without loss of generality.
For MOS evaluation, $20$ Mandarin and $20$ English sentences are randomly selected from the test set to synthesize speech foreign to the target speaker. 
For DMOS and AB tests, we randomly select $10$ Mandarin sentences from the test set to synthesize speech with $6$ types of emotions for two target speakers, respectively, resulting in $120$ testing sentences. 
These synthesized emotional speech sentences also are evaluated for naturalness by MOS.
Twenty Chinese native speakers with basic English skills participated in these experiments. The gender distribution was balanced, and their ages ranged from 20 to 30. The final score for each utterance was the average rating by all participants for this sample.
The results are associated with 95$\%$ confidence intervals in all tests. 
Besides, speaker cosine similarity and embedding visualization are adopted to evaluate speaker similarity and emotion discrimination objectively.

\vspace{-0.2cm}
\subsection{Training setups}
\label{exset}
 
All the speech sentences are down-sampled to $16$ KHz and represented by 80-band Mel-spectrum with a frame length of 50ms, frameshift of 12.5ms, and hop size of $200$.
A grapheme-to-phoneme (G2P) module converts text sentences into phoneme sequences.
The phoneme duration is obtained by a pre-trained Montreal Forced Alignment (MFA) tool~\cite{Ren2019FastSpeechFR}.
We train all the models for 300K iterations with a batch size of 38 on 4 NVIDIA Tesla V100 GPUs.
During the inference, a well-trained Hifi-Gan~\cite{kong2020hifi} is adopted as the neural vocoder to reconstruct waveform from the predicted Mel-spectrum.

\vspace{-0.2cm}
\subsection{Compared methods}
As this work, to our knowledge, is the first time that attempts to synthesize foreign emotional speech based on emotion transfer by a DPM-based model, there is no existing method that can be compared directly. 
Therefore, we selected the most recent relevant methods to compare with our proposed DiCLET-TTS. 
For fairness, some modifications are made to make the compared methods suitable for cross-lingual emotion transfer.
The comparative model and the corresponding improvements are as follows: 
1) \textbf{M3}~\cite{shang2021incorporating} is a FastSpeech-based~\cite{Ren2019FastSpeechFR} multi-speaker multi-style multi-lingual speech synthesis method that introduced a fine-grained style encoder to relieve the foreign accent problem. 
To make M3 suitable for emotion transfer, the emotion ID and emotion classifier is introduced in the style predictor and style encoder, respectively. 
2) \textbf{CSET}~\cite{Li2021ControllableCE} is a reference-based cross-speaker emotion transfer method, which introduced an emotion disentangling module to Tacotron2. 
The text encoder and decoder are extended by the speaker adversarial training and language embedding~\cite{zhang2019learning}, respectively. 
3) \textbf{Grad-TTS}~\cite{popov2021grad} is also improved for cross-lingual emotion transfer.
We follow the original setting of Grad-TTS, where the decoder's input is concatenated with the speaker embedding and emotion embedding obtained from two look-up tables with the speaker ID and emotion ID as input, respectively.
The text encoder structure is the same as that in DiCLET-TTS and is trained by the speaker adversarial loss. 

\section{Experimental results}
\label{sc:results}

In this section, the results of emotions transferred to the intra- and cross-lingual target speakers are presented, i.e., the comparison of DiCLET-TTS with other methods in naturalness, speaker similarity, and emotion similarity.
The corresponding demos can be found on the project page\textsuperscript{\ref{ft:homepage}}, and we recommend readers listen to those demos.

\vspace{-0.2cm}
\subsection{Performance on naturalness}

Two MOS tests are conducted to evaluate the naturalness of Mandarin emotional speech and cross-lingual neutral speech generated by DiCLET-TTS, M3, CSET, and Grad-TTS for intra- and cross-lingual target speakers.
The results are shown in Table~\ref{tab:naturalness}, and unsurprisingly, the highest MOS scores are obtained when synthesizing intra-lingual neutral speech for the target speaker in each method since the synthesized speech is unaffected by the foreign accent and emotion.
DiCLET-TTS achieves higher scores in cross-lingual neutral speeches, while there is no significant difference in scores between the compared methods. 
This may be due to the fact that the text encoder in these three compared methods is only constrained by speaker adversarial training, which could somewhat disturb the linguistic coding.
In DiCLET-TTS, this disturbance is mitigated by the content loss to stabilize the training and effectively improve the naturalness.

\begin{table*}[h]
 \caption{Speaker and emotion similarity DMOS comparison of DiCLET-TTS, M3, CSET, and Grad-TTS in transferring the emotions to intra- and cross-lingual target speakers, with a confidence interval of 95$\%$. The bold indicates the best performance of the four models in each emotion.
 }
 \label{tab:emotiontransfer}
\setlength{\tabcolsep}{3.5mm}
 \centering
\begin{tabular}{l|cccc|cccc}
\toprule
\multicolumn{1}{c|}{\multirow{3}{*}{Emotion}} & \multicolumn{8}{c}{Intra-lingual scenario (target Mandarin speaker)}  \\ \cmidrule{2-9} 
       & \multicolumn{4}{c|}{Speaker similarity DMOS}                                                  & \multicolumn{4}{c}{Emotion similarity DMOS}                                                \\ \cmidrule{2-9} 
       & \multicolumn{1}{c}{M3} & \multicolumn{1}{c}{CSET} & \multicolumn{1}{c}{Grad-TTS} & \multicolumn{1}{c|}{DiCLET-TTS} & \multicolumn{1}{c}{M3} & \multicolumn{1}{c}{CSET} & \multicolumn{1}{c}{Grad-TTS} & \multicolumn{1}{c}{DiCLET-TTS} \\ \midrule
Fear             &\bf{4.04}$\pm$0.04 &3.91$\pm$0.02 &4.02$\pm$0.01 &4.01$\pm$0.05   &3.85$\pm$0.03 &3.71$\pm$0.06 &3.17$\pm$0.03 &\bf{4.04}$\pm$0.04  \\
Disgust          &4.05$\pm$0.02 &3.96$\pm$0.04 &4.06$\pm$0.07 &\bf{4.08}$\pm$0.04   &3.79$\pm$0.05 &3.60$\pm$0.03 &3.13$\pm$0.05 &\bf{3.90}$\pm$0.06    \\ 
Angry            &\bf{4.01}$\pm$0.06 &3.87$\pm$0.03 &3.97$\pm$0.05 &3.98$\pm$0.08   &3.81$\pm$0.06 &3.68$\pm$0.08 &3.19$\pm$0.11 &\bf{3.96}$\pm$0.09    \\
Sadness          &4.02$\pm$0.02 &3.77$\pm$0.05 &\bf{4.03}$\pm$0.02 &4.00$\pm$0.06   &3.90$\pm$0.04 &3.89$\pm$0.04 &3.28$\pm$0.08 &\bf{4.02}$\pm$0.03    \\
Happy            &\bf{3.97}$\pm$0.05 &3.79$\pm$0.04 &3.94$\pm$0.06 &\bf{3.96}$\pm$0.04   &3.92$\pm$0.07 &3.87$\pm$0.06 &3.30$\pm$0.04 &\bf{4.04}$\pm$0.08    \\
Surprise         &3.94$\pm$0.06 &3.84$\pm$0.07 &\bf{4.01}$\pm$0.04 &3.97$\pm$0.07   &3.87$\pm$0.03 &3.82$\pm$0.04 &3.25$\pm$0.07 &\bf{3.97}$\pm$0.06   \\ \midrule
\multicolumn{1}{c}{\multirow{2}{*}{}} &  \multicolumn{8}{c}{Cross-lingual scenario (target English speaker)}  \\ \midrule 
Fear             &\bf{3.91}$\pm$0.05 &3.70$\pm$0.04 &3.86$\pm$0.06 &3.89$\pm$0.02   &3.64$\pm$0.07 &3.55$\pm$0.05 &3.07$\pm$0.09 &\bf{3.86}$\pm$0.06  \\
Disgust          &\bf{3.94}$\pm$0.04 &3.74$\pm$0.07 &3.92$\pm$0.08 &3.90$\pm$0.09   &3.51$\pm$0.08 &3.39$\pm$0.04 &3.05$\pm$0.08 &\bf{3.81}$\pm$0.07   \\ 
Angry            &\bf{3.81}$\pm$0.05 &3.66$\pm$0.10 &3.78$\pm$0.04 &3.79$\pm$0.07   &3.62$\pm$0.11 &3.56$\pm$0.03 &3.14$\pm$0.04 &\bf{3.84}$\pm$0.09   \\
Sadness          &\bf{3.87}$\pm$0.09 &3.64$\pm$0.05 &3.76$\pm$0.03 &3.85$\pm$0.06   &3.57$\pm$0.09 &3.41$\pm$0.08 &3.19$\pm$0.07 &\bf{3.91}$\pm$0.03  \\
Happy            &3.72$\pm$0.04 &3.65$\pm$0.02 &3.75$\pm$0.07 &\bf{3.80}$\pm$0.08   &3.68$\pm$0.04 &3.64$\pm$0.06 &3.21$\pm$0.10 &\bf{3.93}$\pm$0.05   \\
Surprise         &3.68$\pm$0.06 &3.68$\pm$0.09 &3.71$\pm$0.03 &\bf{3.74}$\pm$0.07    &3.60$\pm$0.05 &3.55$\pm$0.02 &3.20$\pm$0.06 &\bf{3.79}$\pm$0.04   \\ \bottomrule 
\end{tabular}
\end{table*}

For synthesized Mandarin emotional speech, generally speaking, the naturalness of transferring emotion to the intra-lingual target speaker is better than transferring emotion to the cross-lingual target speaker among all methods.
This phenomenon is caused by the fact that emotion could make the tone change more violently and the foreign accent more obvious.
The score gap between the synthesized Mandarin emotional speech for intra- and cross-lingual target speakers in DiCLET-TTS and M3 is smaller than that in Grad-TTS and CSET. 
This advantage mainly comes from DiCLET-TTS and M3 adopting prosodic-related linguistic representation, which can alleviate the foreign accent problem and improve the naturalness of cross-lingual emotion transfer.
Besides, DiCLET-TTS achieves the highest naturalness score in synthesized neutral and emotional speeches, indicating that the proposed method can disentangle speakers and languages while stabilizing the training, making the speakers speak foreign languages fluently and express various emotions in authentic Mandarin.

\vspace{-0.4cm}
\subsection{Performance on emotion transfer}
\label{sc:emotrans}

Besides measuring the naturalness, the target speaker similarity and transferred emotion similarity are also evaluated.
Four DMOS tests are conducted to evaluate the speaker similarity and emotion similarity of generated Mandarin emotional speech by DiCLET-TTS, M3, CSET, and Grad-TTS for intra- and cross-lingual target speakers.
The results are shown in Table~\ref{tab:emotiontransfer}, where the upper part is the speaker and emotion similarity results of transferring emotion from the source speaker to the intra-lingual target speaker, the lower part is the results of transferring emotion from the source speaker to the cross-lingual target speaker.

As seen in Table~\ref{tab:emotiontransfer}, regarding the speaker similarity of all emotion categories in each method, the scores of synthesized emotional speech of the intra-lingual target speaker are higher than that of the cross-lingual target speaker. 
This phenomenon could be partially caused by emotion and language affecting participants' perception since the compared reference during the DMOS test of the cross-lingual target speaker is neutral English audio rather than Mandarin emotional audio.
A similar situation also occurs in emotion similarity DMOS.
These results indicate that compared with the cross-speaker emotion transfer task, which only recombines the two factors (speaker, emotion), it is more challenging to simultaneously recombine the three factors (speaker, language, and emotion), which are deeply entangled.

Specifically, regarding speaker similarity, the difference between DiCLET-TTS, M3, and Grad-TTS are not noticeable, while CSET performs the worst.
Although the emotion similarity of CSET is better than Grad-TTS, the poor scores in speaker similarity and cross-lingual naturalness (see Table~\ref{tab:naturalness}) indicate the weakness of CSET for the cross-lingual emotion transfer task.
Grad-TTS achieves reasonable speaker similarity in transferring the emotion to intra- and cross-lingual speakers but performs poorly in emotion similarity. 
It is mainly caused by Grad-TTS adopting a look-up table in emotion modeling, which produces average emotion expressiveness. 
DiCLET-TTS outperforms Grad-TTS in terms of emotion and speaker similarity, showing that such emotion transfer performance is derived not only from the diffusion model but also from the introduced OP-EDM and emotional adaptor.

\begin{table}[t]
 \caption{Speaker cosine similarity of synthesized speech with the cross-lingual target speaker and emotional source speaker, respectively.}
 \label{tab:cosine}
\setlength{\tabcolsep}{1.4mm}
 \centering
\begin{tabular}{c|c|c|c|c|c}
\toprule
\multicolumn{1}{c|}{Speaker} & \multicolumn{1}{c}{Target speaker} & \multicolumn{1}{c}{M3} & \multicolumn{1}{c}{CSET}  & \multicolumn{1}{c}{Grad-TTS} & \multicolumn{1}{c}{DiCLET-TTS} \\ \midrule
    Source speaker &\it{0.18}   &0.23  &0.29  &\bf{0.21}  &0.25 \\\midrule
    Target speaker &\it{0.80}   &\bf{0.75}  &0.65  &0.72  &0.73 \\
  \bottomrule
\end{tabular}
\vspace{-0.3cm}
\end{table}

DiCLET-TTS significantly outperforms all comparison methods in emotion similarity and obtains a comparable speaker similarity score with M3.
The slight speaker similarity gap between M3 and DiCLET-TTS could be caused by the stronger emotional expressiveness of DiCLET-TTS, which could affect participants on the rating of the timbre similarity.
Besides, M3 and DiCLET-TTS adopt speaker adversarial training to remove speaker-related information in emotion embedding. 
The emotional information conveyed by such disentangled emotion embedding tends to be weakened since the speaker and emotion are deeply entangled and both related to the prosody.
While in DiCLET-TTS, the emotion embedding space obtained by OP-EDM is further constrained to ensure that the emotion embedding retains high emotion discrimination after removing the speaker-related information, thus promoting the expressiveness of transferred emotions.
These results show that DiCLET-TTS can well balance maintaining the target speaker's identity and enriching the transferred emotion expressiveness in intra- and cross-lingual scenarios.

\begin{table*}[h]
 \caption{Speaker and emotion similarity DMOS comparison of DiCLET-TTS, ``w/o EA'' and ``w/o CE-D'' in transferring the emotion to the cross-lingual target speaker, with a confidence interval of 95$\%$, and the higher value means better performance and the bold indicates the best performance out of four models in terms of each emotion.
 $\mu$ and $\mu_{emo}$ represent emotion-irrelevant and emotion-related linguistic representation, respectively.}
 \label{tab:abemospk}
\setlength{\tabcolsep}{3mm}
 \centering
\begin{tabular}{l|ccc|ccc}
\toprule 
\multicolumn{1}{c|}{\multirow{2}{*}{Emotion}} & \multicolumn{3}{c|}{Speaker similarity DMOS}   & \multicolumn{3}{c}{Emotion similarity DMOS}                                                \\ \cmidrule{2-7} 
\multicolumn{1}{c|}{}    & \multicolumn{1}{c}{``w/o EA'' ($\mu$)} & \multicolumn{1}{c}{``w/o CE-D''  ($\mu_{emo}$)} & \multicolumn{1}{c|}{DiCLET-TTS ($\mu_{emo}$)} 
                        & \multicolumn{1}{c}{``w/o EA'' ($\mu$)} & \multicolumn{1}{c}{``w/o CE-D''  ($\mu_{emo}$)} & \multicolumn{1}{c}{DiCLET-TTS ($\mu_{emo}$)} \\ \midrule
Fear             &\bf{3.91}$\pm$0.05  &3.83$\pm$0.12   &\bf{3.89}$\pm$0.02      &3.40$\pm$0.09  &3.53$\pm$0.05  &\bf{3.86}$\pm$0.06  \\
Disgust          &\bf{4.00}$\pm$0.04  &3.88$\pm$0.07   &\bf{3.96}$\pm$0.09      &3.41$\pm$0.08  &3.46$\pm$0.04  &\bf{3.81}$\pm$0.07   \\ 
Angry            &\bf{3.82}$\pm$0.03  &3.73$\pm$0.05   &3.79$\pm$0.07           &3.60$\pm$0.04  &3.66$\pm$0.03  &\bf{3.84}$\pm$0.09   \\
Sadness          &\bf{3.90}$\pm$0.06  &3.77$\pm$0.09   &\bf{3.83}$\pm$0.06      &3.43$\pm$0.07  &3.61$\pm$0.08  &\bf{3.91}$\pm$0.03  \\
Happy            &\bf{3.79}$\pm$0.06  &3.66$\pm$0.04   &3.72$\pm$0.08           &3.66$\pm$0.10  &3.72$\pm$0.06  &\bf{3.93}$\pm$0.05   \\
Surprise         &\bf{3.75}$\pm$0.08  &3.63$\pm$0.04   &3.68$\pm$0.07           &3.58$\pm$0.06  &3.69$\pm$0.02  &\bf{3.79}$\pm$0.04   \\ \bottomrule
\end{tabular}
\end{table*}

\vspace{-0.2cm}
\subsection{Speaker similarity with target speaker and source speaker in cross-lingual emotion transfer}

To objectively show the speaker leakage degree of each method, we calculate the speaker cosine similarity between synthesized speech and ground-truth neutral speech from the cross-lingual target speaker and emotional source speaker, respectively. 
Specifically, we adopt a pre-trained speaker verification model ECAPA-TDNN~\cite{TDNN} to extract the x-vectors of synthesized and ground truth speech. 
The speaker cosine similarity with the target speaker and the source speaker has measured on $80$ synthesized speech.

We first calculated the upper bound of cosine similarity within the target speaker's ground truth speech, and the lower bound between the target speaker and the source speaker. 
As shown in Table~\ref{tab:cosine}, the upper bound is \textbf{0.80}, and the lower bound is \textbf{0.18}.
Note that the synthesized speech from CSET has the highest similarity with the source speaker and the lowest similarity with the target speaker, consistent with the results shown in Section~\ref{sc:emotrans}.
The speech synthesized by DiCLET-TTS achieves a comparable cosine similarity score with M3, and as explained above, this gap may also be caused by the stronger emotion expressiveness of DiCLET-TTS. 

\vspace{-0.05cm}
\section{Component analysis}
\label{component}

In Section~\ref{sc:results}, DiCLET-TTS has shown good performance on emotion transfer in intra- and cross-lingual scenarios. 
In this section, the effectiveness of each proposed component, i.e., content loss, emotional adaptor, and condition-enhanced DPM decoder, is evaluated by transferring emotion to the cross-lingual target speaker.
The advantages of the proposed orthogonal projection based emotion disentanglement module (OP-EDM) are also analyzed. 

\vspace{-0.1cm}
\subsection{The effectiveness of content loss and emotional adaptor on naturalness}
\label{abneu}
In DiCLET-TTS, the content loss and emotional adaptor are the keys to improving the naturalness of synthesized cross-lingual speech.
Besides, with the guidance of emotion embedding extracted by OP-EDM, the emotional adaptor and condition-enhanced DPM decoder are further committed to enhancing emotion expressiveness.
Therefore, we first conduct an ablation study via the MOS test to verify the benefits of content loss and emotional adaptor in improving naturalness. 
We do not verify the benefits of the condition-enhanced DPM decoder since it contributes little to improving naturalness.
Specifically, two variants are evaluated: 
1) no content loss is adopted for the text encoder's output, which is constrained only by speaker adversarial training. We denote this variant as ``w/o CTL''. 
2) No emotional adaptor is adopted for the length regulator's output.
We denote this variant as ``w/o EA''.

Table~\ref{tab:abmos} shows the naturalness MOS results of DiCLET-TTS and its two variants. 
Comparing DiCLET-TTS and ``w/o CTL'', we can find the drop of naturalness when discarding content loss in ``w/o CTL'', indicating that introducing content loss in adversarial training can effectively improve the naturalness in synthesized speech.
We also find that the degradation is more prominent in some emotion categories, i.e., \textit{happy, surprise}, and \textit{angry}, since the intonation changes in these categories are more dramatic.
Besides, the naturalness significantly drops in ``w/o EA'', where the linguistic representation is emotion-irrelevant.
This result suggests that parameterizing the terminal distribution of the diffusion process into emotion-related linguistic prior by the emotional adaptor plays an essential role in promoting naturalness.

\begin{table}[htbp]
 \caption{Naturalness MOS results of DiCLET-TTS, ``w/o CTL'', and ``w/o EA'' in transferring emotion to the cross-lingual target speaker, with confidence intervals of 95$\%$. Neutral (Mandarin) and neutral (English) represent synthesized neutral Mandarin and English speech, respectively.}
 \label{tab:abmos}
\setlength{\tabcolsep}{3mm}
 \centering
\begin{tabular}{l|c|c|c}
\toprule
\multicolumn{1}{c|}{Method} & \multicolumn{1}{c}{``w/o CTL''} & \multicolumn{1}{c}{``w/o EA''} & \multicolumn{1}{c}{DiCLET-TTS} \\ \midrule
Neutral (Mandarin)    &3.93$\pm$0.04    &3.86$\pm$0.05    &\bf{3.98}$\pm$0.05   \\
Neutral (English)   &4.15$\pm$0.07         &4.11$\pm$0.04    &\bf{4.21}$\pm$0.05   \\\midrule
Fear                  &3.84$\pm$0.09         &3.71$\pm$0.07    &\bf{3.90}$\pm$0.06   \\
Disgust               &3.85$\pm$0.08         &3.76$\pm$0.12    &\bf{3.93}$\pm$0.04   \\ 
Angry                 &3.71$\pm$0.05         &3.66$\pm$0.11    &\bf{3.82}$\pm$0.07   \\
Sadness               &3.83$\pm$0.10         &3.74$\pm$0.07    &\bf{3.88}$\pm$0.07   \\
Happy                 &3.74$\pm$0.09         &3.70$\pm$0.05    &\bf{3.84}$\pm$0.08    \\
Surprise              &3.72$\pm$0.07         &3.69$\pm$0.08    &\bf{3.83}$\pm$0.05    \\ \bottomrule 
\end{tabular}
\end{table}

\vspace{-0.2cm}
\subsection{The effectiveness of emotional adaptor and condition-enhanced DPM decoder on speaker and emotion similarity}
The effectiveness of the emotional adaptor in improving naturalness has been verified in Section~\ref{abneu}.
In this section, we further present the benefits of the emotional adaptor and condition-enhanced DPM decoder in terms of the speaker and emotion similarity by two DMOS tests.
Therefore, besides the variant ``w/o EA'', the variant ``w/o CE-D'' is also taken into the test, where the emotion embedding and speaker embedding are concatenated with the input of the decoder rather than being added to each ResBlock.

As shown in Table~\ref{tab:abemospk}, regarding the emotion similarity, the two variants in all categories have dropped compared with DiCLET-TTS, and the degradation of ``w/o EA'' is the most significant. 
The lower emotion similarity of ``w/o EA'' brings a weaker impact on the speaker identity of synthesized speech, resulting in a slightly better performance than DiCLET-TTS in speaker similarity.
Specifically, the emotion modeling of ``w/o EA'' is only completed in the condition-enhanced DPM decoder under the condition of the emotion embedding learned by OP-EDM. 
And the linguistic prior of ``w/o EA'' is emotion-irrelevant.
This result indicates that parameterizing the terminal distribution of the diffusion process as an emotion-related linguistic prior by the emotional adaptor can also effectively improve the expressiveness of transferred emotion. 
Besides, referring to the results in Table~\ref{tab:emotiontransfer}, the emotion similarity of ``w/o EA'' is superior to that of Grad-TTS in terms of all emotion categories, and ``w/o EA'' also has an improved performance than CSET in most cases (except \textit{disgust}).
These results also reflect the effectiveness of the introduced OP-EDM in learning speaker-irrelevant emotion embedding, which can result in a good performance in terms of speaker similarity and emotional expressiveness.

For ``w/o CE-D'', although it achieves better performance than ``w/o EA'' on emotion expressiveness,
this improvement is not always significant.  
Emotion expressiveness is still unsatisfactory for emotions (e.g., $disgust$ and $fear$) that rely on speaking speed and stress. 
Meanwhile, for emotions partially reflected in the changes of the source speaker's timbre (e.g., $happy$ and $surprise$), the target speaker similarity of ``w/o CE-D'' is dropped.
All these results show that with the guidance of speaker-irrelevant emotion embedding extracted from OP-EDM, the emotional adaptor and condition-enhanced DPM decoder can effectively improve the performance of cross-lingual emotion transfer while maintaining reasonable speaker similarity and speech naturalness.

\vspace{-0.2cm}
\subsection{Advantages of emotion embedding space with orthogonal projection}

This section analyzes the benefits of the proposed orthogonal projection based emotion disentanglement module (OP-EDM) by comparing it with the variant ``w/o OPL'', in which ``w/o OPL'' means the orthogonal projection loss in OP-EDM is removed.
Ideally, the emotion embedding learned by the emotion disentanglement module is expected to be irrelevant to the speaker identities but holds high emotion discrimination.
Therefore, the t-distributed stochastic neighbor embedding (t-SNE)~\cite{Laurens2008Visualizing} is adopted to demonstrate the capacity of emotion embedding learned from these two modules on distinguishing emotion categories or speaker identities.

\begin{figure}[t]
	\centering 
	\begin{minipage}{\linewidth}
		\begin{minipage}{0.49\linewidth}
			\includegraphics[width=\textwidth]{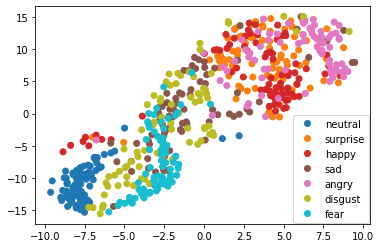}
			\centerline{(a) ``w/o OPL''}
		\end{minipage}
		\hfill
		\begin{minipage}{0.49\linewidth}
			\includegraphics[width=\textwidth]{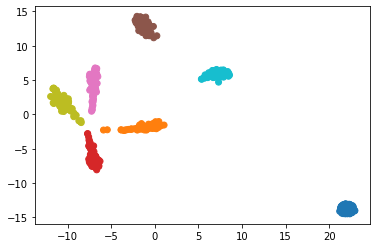}
			\centerline{(b) OP-EDM}
		\end{minipage}
		\vfill
	\end{minipage}	
	\caption{Emotion distribution of the emotion embedding created different models (a) ``w/o OPL'' and (b) OP-EDM. The presented data are $80$ sentences randomly selected from each emotion category of the \textit{CN-emo}'s test set.}
	\label{fig:emb_tsne}
 \vspace{-0.15cm}
\end{figure}
\begin{figure}[t]
	\centering
	\begin{minipage}{\linewidth}
		\begin{minipage}{0.49\linewidth}
			\includegraphics[width=\textwidth]{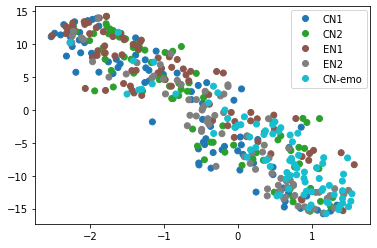}
			\centerline{(a) ``w/o OPL''}
		\end{minipage}
		\hfill
		\begin{minipage}{0.49\linewidth}
			\includegraphics[width=\textwidth]{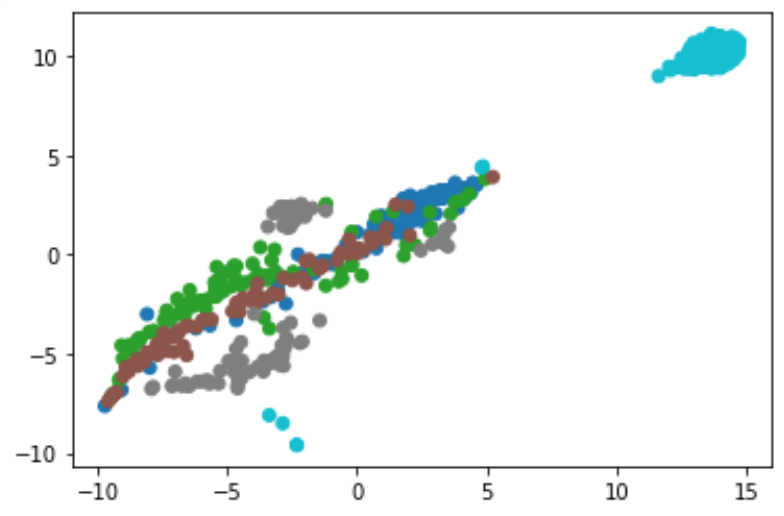}
			\centerline{(b) OP-EDM}
		\end{minipage}
		\vfill
	\end{minipage}
	\caption{Speaker distribution of the emotion embedding created by different models (a) ``w/o OPL'' and (b) OP-EDM module. The presented data are $80$ neutral sentences randomly selected from each speaker's test set.}
	\label{fig:spk_tsne}
\vspace{-0.1cm} 
\end{figure}

\subsubsection{Emotion discrimination ability}

To display the distribution of emotion embeddings extracted by ``w/o OPL'' and OP-EDM, $80$ speeches of each emotion category from the \textit{CN-emo}'s test set are randomly selected, resulting in $560$ reference speeches in total and then embedded as emotion embeddings by these two modules, respectively.
The distributions of these embeddings are presented in Fig.~\ref{fig:emb_tsne}, where each point indicates an emotion embedding, and points with the same color are from the same emotion category. 
Smaller distances between the two points indicate that the embeddings are more similar.
As shown in Fig.~\ref{fig:emb_tsne} (a), the emotional embedding generated by ``w/o OPL'' only retains weak emotion discrimination, where emotions with similar characteristics tend to be confused: (1) \textit{happy}, \textit{surprise}, and \textit{angry} with a higher pitch and fast speech speed; (2) \textit{sad}, \textit{fear}, and \textit{disgust} with a deep voice and slower speech speed.
In contrast, in Fig.~\ref{fig:emb_tsne} (b), the emotion embeddings from the same emotion category are clustered together while different clusters are separated, demonstrating that benefits from the orthogonal projection loss, OP-EDM can obtain an embedding space with high emotion discrimination. 
We notice that although these embeddings are all from the same speaker, the neutral embeddings are far away from the others. This phenomenon could be due to the fact that emotions are mainly reflected in pitch, energy, and speech speed, and these attributes are relatively flat in neutral emotions.

\subsubsection{Speaker identity removal capability}
For speaker identity visualization, $80$ neutral speeches are randomly selected from each speaker's test set, resulting in $400$ speeches. 
The visualization results are shown in Fig.~\ref{fig:spk_tsne}, in which the same color colors the embeddings from the identical speaker.
As mentioned, the emotion embedding should contain no speaker-related information but only emotional information, which implies embeddings extracted from different speakers' speech are expected to be inseparable.
As shown in Fig.~\ref{fig:spk_tsne}(a), the embeddings from different speakers extracted by ``w/o OPL'' are indeed inseparable, while the cost is that these embeddings maintain little emotion information from the reference audio (see Fig.~\ref{fig:emb_tsne}(a)).
For the OP-EDM module (see Fig.~\ref{fig:spk_tsne}(b)), the embeddings from the four neutral speakers' corpus are clustered into one cluster. 
It is worth noting that the neutral speech from \textit{CN-emo} is treated as an independent emotion category, so the embeddings from \textit{CN-emo} are clustered into a separate cluster in Fig.~\ref{fig:spk_tsne}(b).
This distribution indicates that the proposed OPL-EDM can effectively remove the speaker-related information while greatly retaining the emotion-related information, resulting in speaker-irrelevant but emotion-discriminative embedding. 

\begin{figure}[t]
\begin{minipage}[b]{1.0\linewidth}
  \centering
  \centerline{\includegraphics[height=5cm,width=8.5cm]{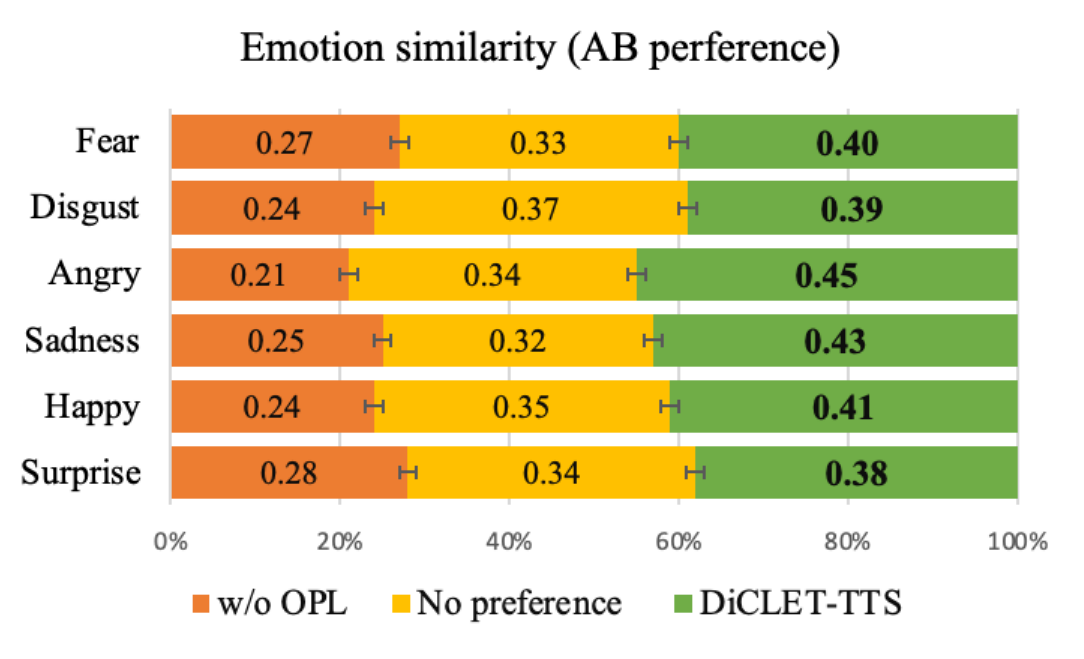}}
 \vspace{-0.2cm}
  \caption{Emotion similarity AB preference test for ``w/o OPL'' and DiCLET-TTS with confidence intervals of 95\%.}
  \label{fig:abemo}
 \end{minipage}
\vspace{-0.2cm} 
\end{figure}

\begin{figure}[t]
\begin{minipage}[b]{1.0\linewidth}
  \centering
  \centerline{\includegraphics[height=5cm,width=8.5cm]{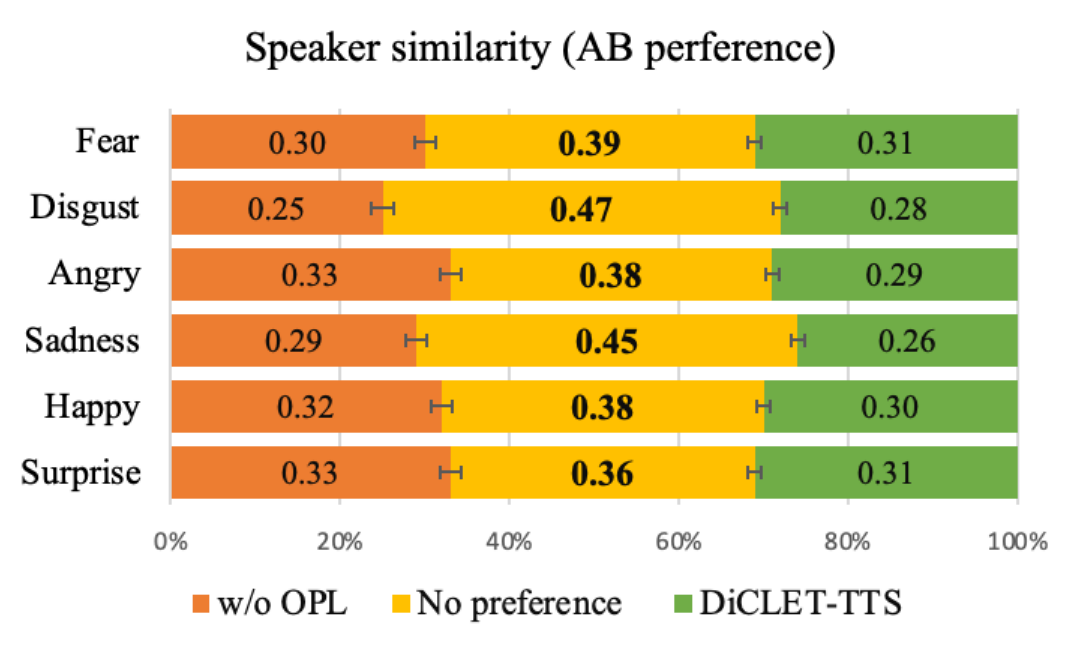}}
 \vspace{-0.2cm}
  \caption{Speaker similarity AB preference test for ``w/o OPL'' and DiCLET-TTS with confidence intervals of 95\%.}
  \label{fig:abspk}
 \end{minipage}
\end{figure}

\subsubsection{Preference test}
To further investigate the effectiveness of using OPL in learning emotion embedding for emotion transfer.
We conducted two AB tests between DiCLET-TTS and the variant ``w/o OPL'' regarding emotion and speaker similarity.
The results are shown in Fig.~\ref{fig:abemo} and Fig.~\ref{fig:abspk}, respectively.
As shown in Fig.~\ref{fig:abemo}, we can find that ``w/o OPL'' obtains lower preference in all emotion categories, showing lower emotion similarity is perceived.
In contrast, the listeners preferred DiCLET-TTS more when we inserted OPL into OP-EDM. 
As analyzed, the performance gain is essentially contributed by the OPL strategy in learning discriminative emotion embeddings.
Regarding speaker similarity, as shown in Fig.~\ref{fig:abspk}, there is no significant difference between ``w/o OPL'' and DiCLET-TTS, i.e., most listeners give \textit{No preference}. 
All the above evidence shows that OP-EDM introduced in this paper contributes to better emotion similarity without reducing speaker similarity.

\section{Conclusion}
\label{sc:conclusion}

This paper proposes a DPM-based cross-lingual emotion transfer model -- DiCLET-TTS.
We adopt prosodic information to alleviate the foreign accent problem, where a prior text encoder takes emotion embedding as a condition to parameterize the terminal distribution of the forward diffusion processes into a speaker-irrelevant but emotion-related linguistic prior.
To address the weaker emotional expressiveness problem caused by removing speaker information from emotion embedding, an orthogonal projection based emotion disentangling module (OP-EDM) is proposed to learn the speaker-irrelevant but high emotion-discriminative embedding.
The reverse diffusion process is parameterized by a condition-enhanced DPM decoder, where the modeling ability of the speaker and emotion is enhanced to further improve emotion expressiveness in synthetic speech.
Experimental results demonstrate that DiCLET-TTS performs well in intra- and cross-lingual emotion transfer while preserving the timbre of the target speaker and synthesized naturalness. 
The results also prove the advantages of OP-EDM in learning speaker-irrelevant but emotion-discriminative embedding.

In this study, only the same-gender speakers are involved in our experiments while cross-gender emotion transfer is considered a difficult task itself and it can be more challenging in the cross-lingual scenario. We will further study this cross-gender task as a follow-up work.

\bibliographystyle{IEEEtran}
\bibliography{mybibfile.bib}
%




\end{document}